\documentclass[twoside,fleqn]{article}
\usepackage{epsf}
\usepackage{rotate}
\usepackage{espcrc2}

\newcommand{\AmS}{{\protect\the\textfont2
  A\kern-.1667em\lower.5ex\hbox{M}\kern-.125emS}}
\newcommand{\be}[1]{\begin{equation} \label{(#1)}}
\newcommand{\ee}{\end{equation}}
\newcommand{\ba}[1]{\begin{eqnarray} \label{(#1)}}
\newcommand{\ea}{\end{eqnarray}}
\newcommand{\nn}{\nonumber}
\newcommand{\rf}[1]{(\ref{(#1)})}
\def\rp{$R_p \hspace{-1em}/\;\:$ }

\def\rpm{R_p \hspace{-0.8em}/\;\:}

\def \znbb {\beta\beta_{0\nu}}

\def \emass {\langle m_{\nu} \rangle}

\title{Neutrinoless double beta decay, neutrino mass and 
bilinear R-parity breaking supersymmetry}

\author{M. Hirsch\address{Instituto de F\'{\i}sica Corpuscular 
-- C.S.I.C. \\ Departamento de F\'{\i}sica Te\`orica, 
Universitat of Val\`encia, \\ Edificio Institutos de Paterna,
Apartado de Correos  2085 \\ 46071 Val\`encia, 
Spain\\ mahirsch@flamenco.ific.uv.es}}

\begin{document}

\begin{abstract}
Neutrinoless double beta ($\znbb$) decay violates lepton number; its 
absence stringently constrains the parameters of theories beyond the 
standard model in which the neutrino has a Majorana mass. R-parity 
violating weak-scale supersymmetry is a prominent example of such 
models. Double beta decay in supersymmetry with explicit bilinear 
R-parity breaking is discussed and current limits on the $\znbb$ decay 
half life of $^{76}$Ge are used to extract upper bounds on the 
R-parity breaking parameters of the first generation. Moreover, 
it is shown that the effective Majorana neutrino mass, measured in 
$\znbb$ decay, is non-zero once the 1-loop corrections 
are taken into account even for the case of perfect alignment 
($\Lambda_i:=(\langle {\tilde \nu}_i \rangle \mu - v_1 \epsilon_i) 
\equiv 0$) among the R-parity violating parameters. 
\end{abstract}

\maketitle

\section{Introduction}

Neutrinoless double beta ($\znbb$) decay is a $\Delta L=2$ process and
therefore one naturally expects it to occur in models with lepton number
violation in the Lagrangian. Even though there is a variety of
mechanisms inducing $\znbb$ decay in gauge theories, one can show
that whatever the leading mechanism is at least one of the neutrinos
will be a Majorana particle~\cite{SV82}, as illustrated in the
black-box diagram of Fig.~1.  This well-known argument establishes a
deep connection between Majorana neutrino masses and $\znbb$ decay: in
gauge theories one can not occur without the other being present. The
same remains true in supersymmetric theories \cite{theorem}, where
moreover one can show that also the supersymmetric partner of the
neutrino must have a $B-L$ violating {\sl Majorana-like} mass term, if
a Majorana mass of the neutrino exists \cite{theorem}.  Turning the
argument around, one expects that the observed absence of $\znbb$
decay allows to derive stringent limits on \rp parameters. This has
been shown for models with explicit trilinear R-parity breaking in
\cite{hir9598,fae97} and for SUSY with explicit bilinear R-parity 
breaking in \cite{hir99,FKS98}. Here, I will mainly report on the 
results derived in \cite{hir99}.

Section 2 sets up the notation and definitions of the bilinear R-parity 
breaking model, whereas section 3 discusses $\znbb$ decay in 
bilinear \rp SUSY at tree level. In section 4, the $\znbb$ decay 
in bilinear \rp SUSY is discussed at the level of 1-loop 
for the first time in literature. 

%
%
\bigskip
\hskip0mm
\epsfysize=45mm
\epsfxsize=70mm
\epsfbox{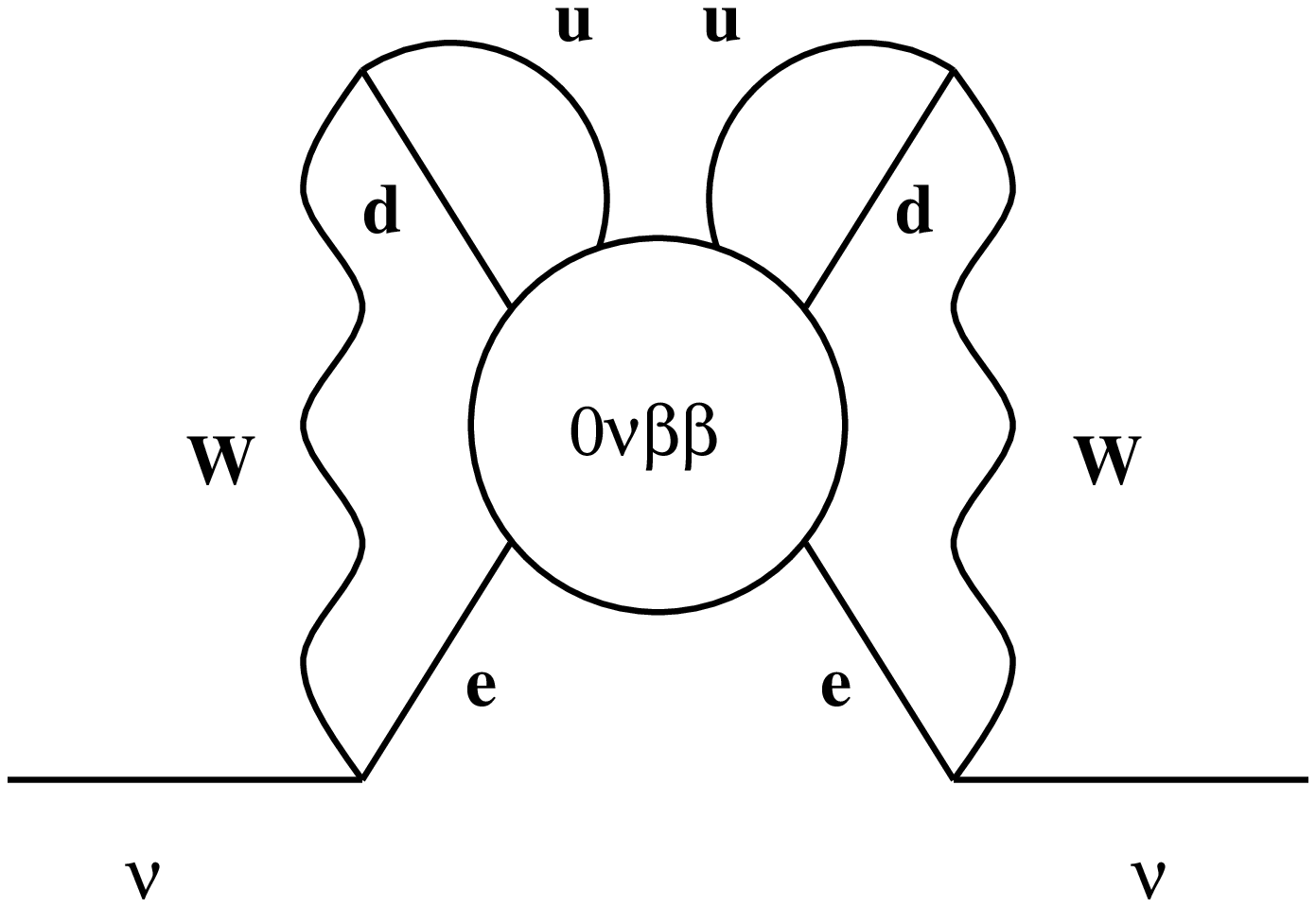}

\noindent
{\bf Figure 1: }{Diagram illustrating the connection between the 
Majorana mass of the neutrino and the amplitude of double beta decay.}

\section{Minimal R-parity broken supersymmetry}

The minimal supersymmetric extension of the standard model (MSSM) has a 
conserved R-parity. This is a multiplicative quantum number which can 
be defined as $R_P = (-1)^{3B+L+2S}$, where $B$ and $L$ are the baryon 
and lepton number and $S$ the spin of the corresponding particle. This 
property of the MSSM \cite{HK85}  is theoretically {\sl ad hoc} since 
the origin of R-parity conservation is unknown. 

In supersymmetry there is actually no distinction between the 
lepton doublet and the Higgs doublet superfield giving mass to 
the down-type quarks unless conservation of lepton number is {\it assumed}. 
This fact can be accounted for by defining a superfield ${\hat \Phi}$ 
as

\be{defphi}
{\hat \Phi} = ({\hat H_1},{\hat L_1},{\hat L_2},{\hat L_3}).
\ee
For the MSSM field content the most general gauge invariant 
form of the renormalizable superpotential can then be written 
as 

\ba{superpotphi}\nn
W &=& \epsilon_{ab} \Big[ 
    \lambda_{e}^{IJk} {\hat \Phi}_I^a {\hat \Phi}_J^b {\hat E}_k^C
  + \lambda_{d}^{Ijk} {\hat \Phi}_I^a {\hat Q}_j^b {\hat D}_k^C \\
  &+&  h_{u}^{jk} {\hat Q}_j^a {\hat H}_2^b {\hat U}_k^C
  +  \mu^I {\hat \Phi}_I^a {\hat H}_2^b \Big] .
\ea
Here, ${\hat Q}$ and ${\hat D}^C$, ${\hat U}^C$ are the quark doublet
and singlets superfields, respectively, ${\hat E}^C$ is the lepton
singlet superfield and ${\hat H}_2$ the Higgs superfields with
$Y({\hat H}_2) = 1$ responsible for the up-type quark masses, with
$h_{u}^{jk}$ being the corresponding Yukawa couplings.  The indices
$j,k~=~1,2,3$ denote generations, whereas $I,J~=~0,1,2,3$. The indices
$a,b$ are $SU(2)$ indices.  The part of $W$ with indices $I,J=0$ 
corresponds to the MSSM superpotential, wheras $I,J=1,2,3$ represents 
the R-parity violating terms. The vector $\mu^I$ is 
$\mu^I =(\mu,\epsilon_e,\epsilon_{\mu},\epsilon_{\tau})$.

If the only source of RPV in the model was found in the 
superpotential one could easily rotate the field ${\hat \Phi}$ into a 
basis ${\hat \Phi}'$ with $\mu^{I'}=(\mu',0,0,0)$ effectively eliminating
the bilinear terms. However, another source of RPV is found in the
soft supersymmetry breaking part of the scalar potential. It contains
the terms:
\ba{vsoftrp} 
V^{soft}_{\rpm} & = & 
{\tilde A}_{e}^{IJk}{\tilde{\Phi}}_I {\tilde {\Phi}}_J {\tilde E^C}_k + 
{\tilde A}_{d}^{Ijk}{\tilde {\Phi}}_I {\tilde Q}_j {\tilde D^C}_k \\ \nn
&+& B^I {\tilde {\Phi}}_I  H_2 
+ (m_{IJ}^2 + \mu_I\mu_J) {\tilde \Phi}_I {\tilde \Phi}_J^{\dagger} 
+ \cdots
\ea
where the dots represent terms not interesting for the discussion 
here. Rotating the superpotential as discussed above, it is easy 
to see that as long as the $B^I$ are not exactly parallel to the 
$\mu^I$ the effects of the bilinear terms of the superpotential 
in the rotated basis will reappear in the soft SUSY breaking terms.

The presence of the bilinear terms in \rf{vsoftrp} imply that in 
general the sneutrino fields acquire Vevs and as a result the 
leptons and gauginos of the model mix. The neutralino mass matrix 
which in the MSSM is a ($4\times 4$) matrix in the bilinear \rp 
MSSM is a ($7\times 7$) matrix, including in addition the three 
generations of neutrinos. It can be written in the following form:

\be{nmm}
{\cal M}_0 =  \left(
                    \begin{array}{cc}
                    0 & m \\
                    m^T & {\cal M}_{\chi^0} \\
                    \end{array}
              \right).
\ee
Here, the submatrix $m$ contains entries from the bilinear \rp 
parameters,

\be{bnmm}
m =   \left(
            \begin{array}{cccc}
     -\frac{1}{2}g'\omega_e & \frac{1}{2}g\omega_e & 0 & -\epsilon_e \\
-\frac{1}{2}g'\omega_{\mu} & \frac{1}{2}g\omega_{\mu} & 0 & -\epsilon_{\mu} \\
        -\frac{1}{2}g'\omega_{\tau} & \frac{1}{2}g\omega_{\tau} & 
          0 & -\epsilon_{\tau} \\
                    \end{array}
              \right),
\ee
$\omega_i := \langle {\tilde \nu}_{i} \rangle$. ${\cal M}_{\chi^0}$ 
is the MSSM neutralino mass matrix given by,

\ba{MSSMnm}  \nn
{\cal M}_{\chi^0} =  \left(
                        \begin{array}{cccc}
 M_1 & 0   & -\frac{1}{2}g' v_1 &  \frac{1}{2} g' v_2  \\
 0   & M_2 &  \frac{1}{2}g  v_1 & -\frac{1}{2} g  v_2  \\
  -\frac{1}{2} g' v_1 &  \frac{1}{2} g v_1 & 0 & -\mu  \\
   \frac{1}{2} g' v_2 & -\frac{1}{2} g v_2 & -\mu & 0 \\
 \end{array}
                     \right)
\ea

It is interesting to note that the matrix \rf{nmm} has such a 
texture that at {\it tree level} only one of the three neutrinos 
gets massive, leaving two massless but mixed states in the 
spectrum. 

If the \rp parameters are small in the sense that for 
\be{exp0}
\xi = m \cdot {\cal M}_{\chi^0}^{-1}
\ee
all $\xi_{ij} \ll 1$, one can find an approximate solution for 
the neutrino/neutralino mass matrix. 

In leading order in $\xi$ the mixing matrix $\Xi$ which 
diagonalizes the mass matrix is given by,
\ba{xiapprox} \nn
\Xi^* & = & \left(\begin{array}{cc}
V_\nu^T & 0 \\
0 & N^* \end{array}\right)
\left(\begin{array}{cc}
1 -{1 \over 2}\xi \xi^{\dagger} & -\xi \\
\xi^{\dagger} &  1 -{1 \over 2}\xi^\dagger \xi
\end{array}\right) \\ 
& = &  \left(\begin{array}{cc}
V_\nu^T(1 -{1 \over 2}\xi \xi^{\dagger} ) & -V_\nu^T\xi \\
N^* \xi^{\dagger} & N^* ( 1 -{1 \over 2}\xi^\dagger \xi) 
\end{array}\right)
\ea
The second matrix in eq. \rf{xiapprox} above block-diagonalizes 
${\cal M}_0$ approximately to the form 
diag($m_{eff},{\cal M}_{\chi^0}$), where

\ba{meff}
&&m_{eff} = - m \cdot {\cal M}_{\chi^0}^{-1} m^T \\ \nn
&=& \frac{M_1 g^2 + M_2 {g'}^2}{4 det({\cal M}_{\chi^0})} 
\left(\begin{array}{ccc}
\Lambda_e^2 & \Lambda_e \Lambda_\mu
& \Lambda_e \Lambda_\tau \\
\Lambda_e \Lambda_\mu & \Lambda_\mu^2
& \Lambda_\mu \Lambda_\tau \\
\Lambda_e \Lambda_\tau & \Lambda_\mu \Lambda_\tau & \Lambda_\tau^2
\end{array}\right)
\ea
Here, $det({\cal M}_{\chi^0})$ is the determinant of ${\cal M}_{\chi^0}$ 
and 
 \be{deflam}
\Lambda_i = \mu \omega_i - v_1 \epsilon_i.
\ee
${\vec \Lambda} := (\Lambda_e,\Lambda_{\mu},\Lambda_{\tau})$, which 
will be called the alignment vector, plays a very prominent role 
in the bilinear \rp model. As one can read off from eq. \rf{meff} 
at tree level neutrinos are massless if ${\vec \Lambda} \equiv 0$. 
Second, even though two neutrinos are massless at tree level, all 
neutrinos mix with each other and the heavy states, except if (and 
only if) the corresponding component $\Lambda_i$ vanishes. 

\section{$\znbb$ decay in the bilinear \rp MSSM at tree level}

In the bilinear \rp MSSM there are 4 Feynman diagrams in lowest 
order of perturbation, see fig. 2. One can show, however, that  
the simplest of these graphs, the neutrino mass contribution 
is always dominant \cite{hir99}. 

The double beta decay observable, under the assumption that 
the \rp parameters are small, can be written as 

\be{defmeff}
\langle m_{\nu} \rangle = \sum_j' U_{ej}^2 m_j 
= \frac{2}{3} \frac{g^2 M_2}{det({\cal M}_{\chi^0})}\Lambda_e^2
\ee
where $\Lambda_e =\omega_e \mu - v_1 \epsilon_e $ and
the prime indicates summation over only light fermion states 
Note, that eq. \rf{defmeff} is proportional 
to the alignment factor $\Lambda_e$, i.e. at tree level double 
beta decay would vanish, as is the case for the neutrino mass, 
in the limit of ${\vec \Lambda} \equiv 0$. Using the experimentally 
measured lower limit on the half life of $\znbb$ decay, Eq. \rf{defmeff} 
can then be used to establish limits on $\Lambda_e$ as a function 
of the R-parity conserving SUSY parameters $\mu$, $M_2$ and 
$\tan\beta$. An example is shown in fig. 3. The small region(s) extending 
to the upper right are the alignment regions were $\Lambda_e 
\rightarrow 0$. Along these lines there is no constraint on 
($\epsilon_e,\omega_e$) from $\znbb$ decay at tree-level.

\vskip-5mm
\hskip-15mm
\epsfysize=100mm
\epsfxsize=85mm
\epsfbox{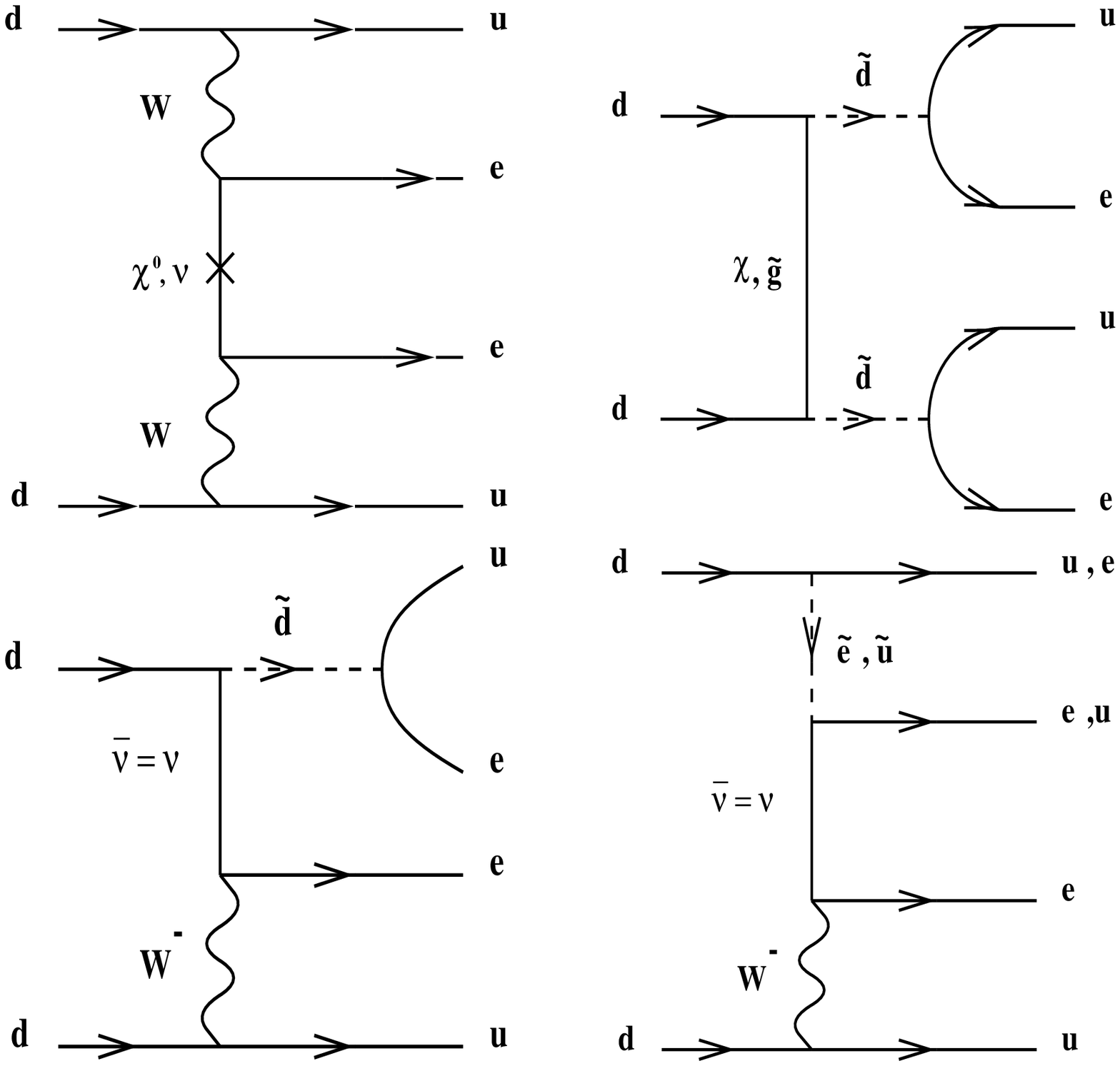}

\vskip-20mm
{\bf Figure 2:}{ Leading order Feynman diagrams in the bilinear 
\rp model.}

\vskip-10mm
\hskip0mm
\epsfysize=70mm
\epsfxsize=70mm
\epsfbox{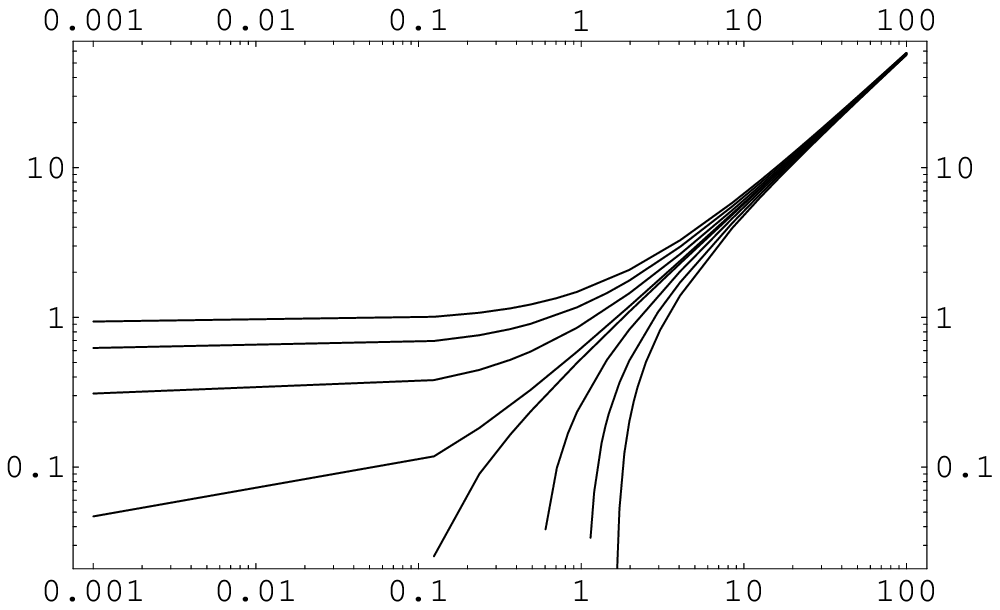}

\vskip-50mm
\noindent
{\rotate{$\epsilon_e$ [MeV]}

\vskip23mm
\hskip50mm
$\omega_e$  [MeV]

\medskip
{\bf Figure 3: }{Excluded ranges in plane ($\epsilon_e,\omega_e$) using 
$\emass \le 0.5$ eV for $\tan\beta = 1$ and $\mu = 100$ GeV, for  
different values of $M_2$, $M_2= 100, 200, 500,
1000$ GeV.  Note that the allowed range is always in between two 
lines of constant $M_2$. }

\bigskip

Another way of visualizing the constraints is to invert the procedure 
and calculate the theoretically expected half lives as a function of, 
for example $\omega_e$. The results of such a study is shown in figure 
4. The experimental limit is taken from \cite{hdmo}. 
Values of $\omega_e$ larger than about $2$ MeV are not allowed, as 
long as $\omega_e$ and $\epsilon_e$ are not perfectly aligned.

%
%
\vskip-10mm
\hskip-8mm
\epsfysize=70mm
\epsfxsize=70mm
\hskip10mm
\epsfbox{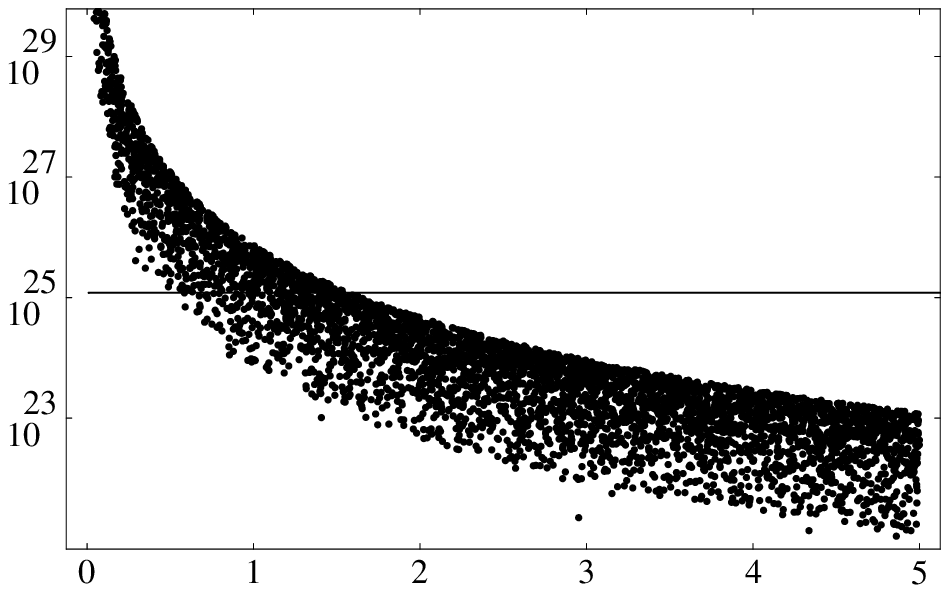}

\vskip-50mm
\noindent
{\rotate{$T_{1/2}^{\znbb}$ [ys]}

\vskip23mm
\hskip50mm
$\omega_e$  [MeV]

%
%
\bigskip
\noindent
{\bf Figure 4: }{ Calculated half-life for the $\znbb$ decay 
of $^{76}Ge$ as function of $\omega_e$ for a random variation 
of the MSSM parameters, $M_2$ and $\mu$ from $100$ GeV to $1$ TeV 
and $\tan\beta=1-50$.}

\bigskip

\section{Some results for $\znbb$ decay at 1-loop}

Since double beta decay in the bilinear model in the lowest 
order of perturbation is strictly proportional to $\Lambda_e$, 
it is an interesting question to ask, whether it is possible 
to determine an absolute upper bound on individual parameters, 
$\epsilon_e$ (and $\omega_e$) even for the special case 
$\Lambda_e \equiv 0$, going to higher orders of 
perturbation. 

There are three simple topologies of relevant Feynman diagrams
contributing to the neutrino-neutralino mass matrix at 1-loop 
\cite{numasses,Paper2}.
With these the one-loop corrected mass matrix is calculated as,
\ba{1loopMass}\nn
M_{ij}^{pole} & =& M_{ij}^{\overline {DR}}(\mu_R) 
              + \frac{1}{2} \Big(
                \Pi_{ij}(p_i^2) + \Pi_{ij}(p_j^2) \\
              &  - & m_{\chi^0_i} \Sigma_{ij}(p_i^2) - 
                  m_{\chi^0_j} \Sigma_{ij}(p_j^2) \Big)
\ea
where $\Sigma_{ij}$ and $\Pi_{ij}$ are self-energies.  For a complete
description see~\cite{Paper2}. Here, ${\overline {DR}}$ signifies the
minimal dimensional reduction subtraction scheme and $\mu_R$ is the
renormalization scale.  

Although the 1-loop corrections for the heavy states (``neutralinos'') 
can be expected to be moderate, for the two lightest states they 
are essential. Moreover, the mixing of the neutrinos to the 
down type higgsinos and to the charged higgsino contains a piece 
which is directly proportional to $\epsilon_i$ and {\it not} only to 
$\Lambda_i$ \cite{hir99,Paper2}. One therefore has to expect, 
that after inclusion of the 1-loop correction $\emass$ will be 
sensitive to the value of $\epsilon_e$ directly. 

Following the procedure of \cite{numasses,Paper2} I have calculated values 
for the 1-loop neutrino mass matrix and deduced  $\emass$ 
for several $10^3$ randomly generated points in SUSY parameter 
space. Results are shown in fig. 5. The ranges of parameters 
were chosen as follows. For the MSSM parameters: $|\mu|$, $M_2 \le 500$ 
GeV, the common scalar mass $m_0 = 0.2-1$ TeV, $\tan\beta = 2.5-20$. 
Since double beta decay is sensitive only to the first generation 
\rp parameters, it suffices to state that in generating the plot 
I have always kept $\Lambda_e$ at least a factor of $100$ smaller 
than required by the tree level bound. The last condition is applied 
to guarantee that only the highly aligned part of parameter space is 
explored.

%
%
\vskip-20mm
\hskip-10mm
\epsfysize=100mm
\epsfxsize=70mm
\hskip10mm
\epsfbox{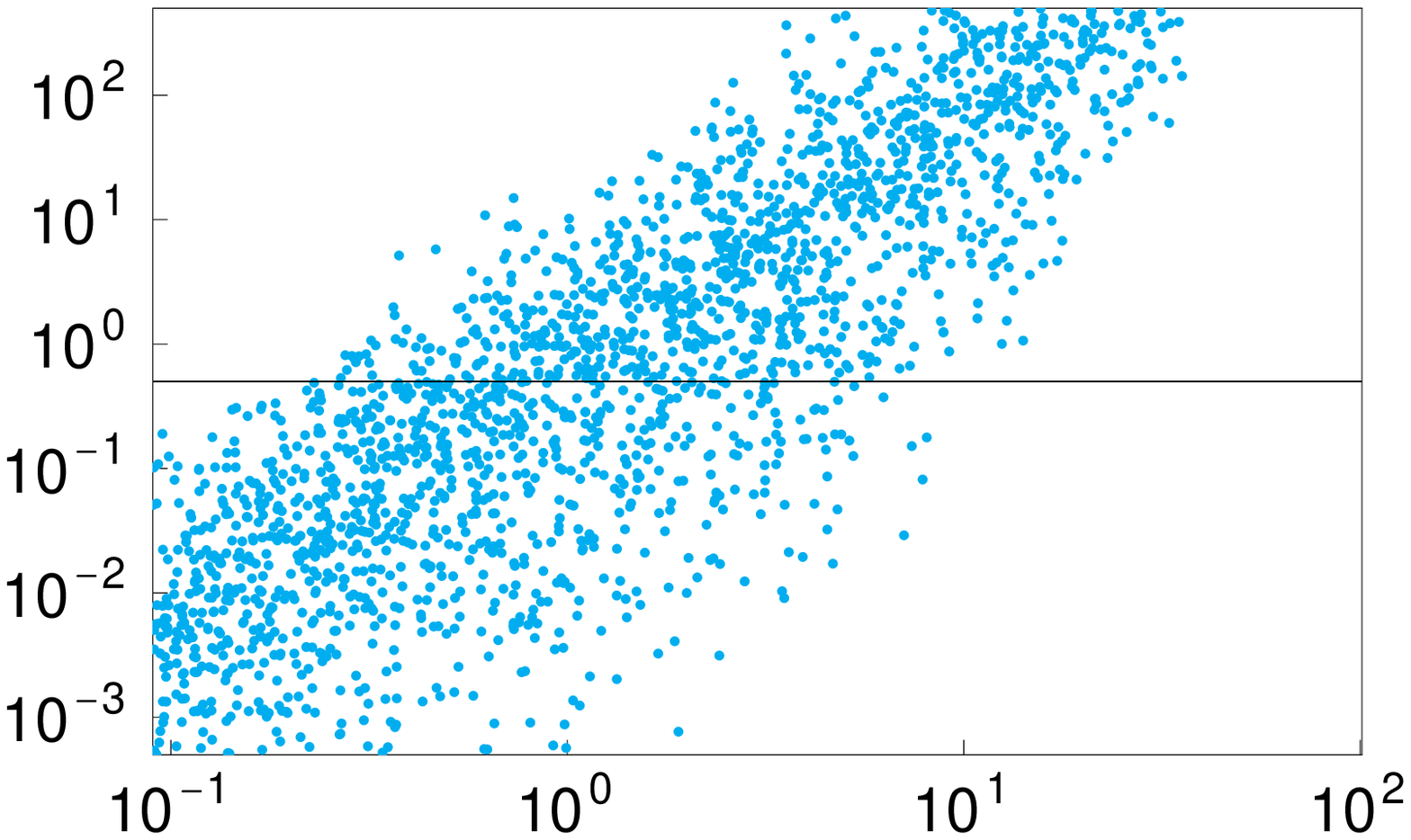
}

\vskip-70mm
\noindent
{\rotate{$\emass$}

\vskip33mm
\hskip50mm
$\epsilon_e$  [GeV]

\bigskip
\noindent
{\bf Figure 5: }{Calculated values of $\emass$ including the 1-loop 
corrections, as a function of $\epsilon_e$, for 
$\Lambda_e \le 10^{-2} \times \Lambda_e^{tree,max}$. 
$\Lambda_e^{tree,max}$ is the maximal $\Lambda_e$ allowed from the 
tree-level analysis. The horizontal line 
indicates the current experimental upper limit.}
 
\bigskip
Because of the complexity of the 1-loop calculation it is impossible 
to derive a semi-analytical bound as in the tree level case. 
Nevertheless, figure 5 shows that $\epsilon_e$ can not be larger 
than about $5-10$ GeV. Note, that larger values of $m_0$ would allow 
only for very slightly larger values of $\epsilon_e$, whereas larger 
values for $\tan\beta$ would give lower upper limits.

It is also interesting to ask how reliable the tree-level estimation 
for the double beta decay observable is. Following the arguments 
about the alignment, discussed above, it is expected that if 
$\epsilon_e^2/\Lambda_e$ is small, i.e. the suppression of 
$\Lambda_e$ is due to the smallness of $\epsilon_e$ and $\omega_e$ 
themselves and not due to cancellations among the two terms in 
$\Lambda_e$ , the tree level expression should be a good 
approximation, wheras for large $\epsilon_e^2/\Lambda_e$ one expects 
the tree level estimate to fail badly. This is demonstrated in fig. 6, 
which nicely confirms this qualitative expectation. Note, however, 
that even for $\epsilon_e^2/\Lambda_e$ as small as (few) $10^{-2}$ 
the tree-level expression is not reliable and it certainly 
fails if $\epsilon_e^2/\Lambda_e$ is larger than order ${\cal O}(1)$.

%
%
\vskip-20mm
\hskip-8mm
\epsfysize=100mm
\epsfxsize=70mm
\hskip10mm
\epsfbox{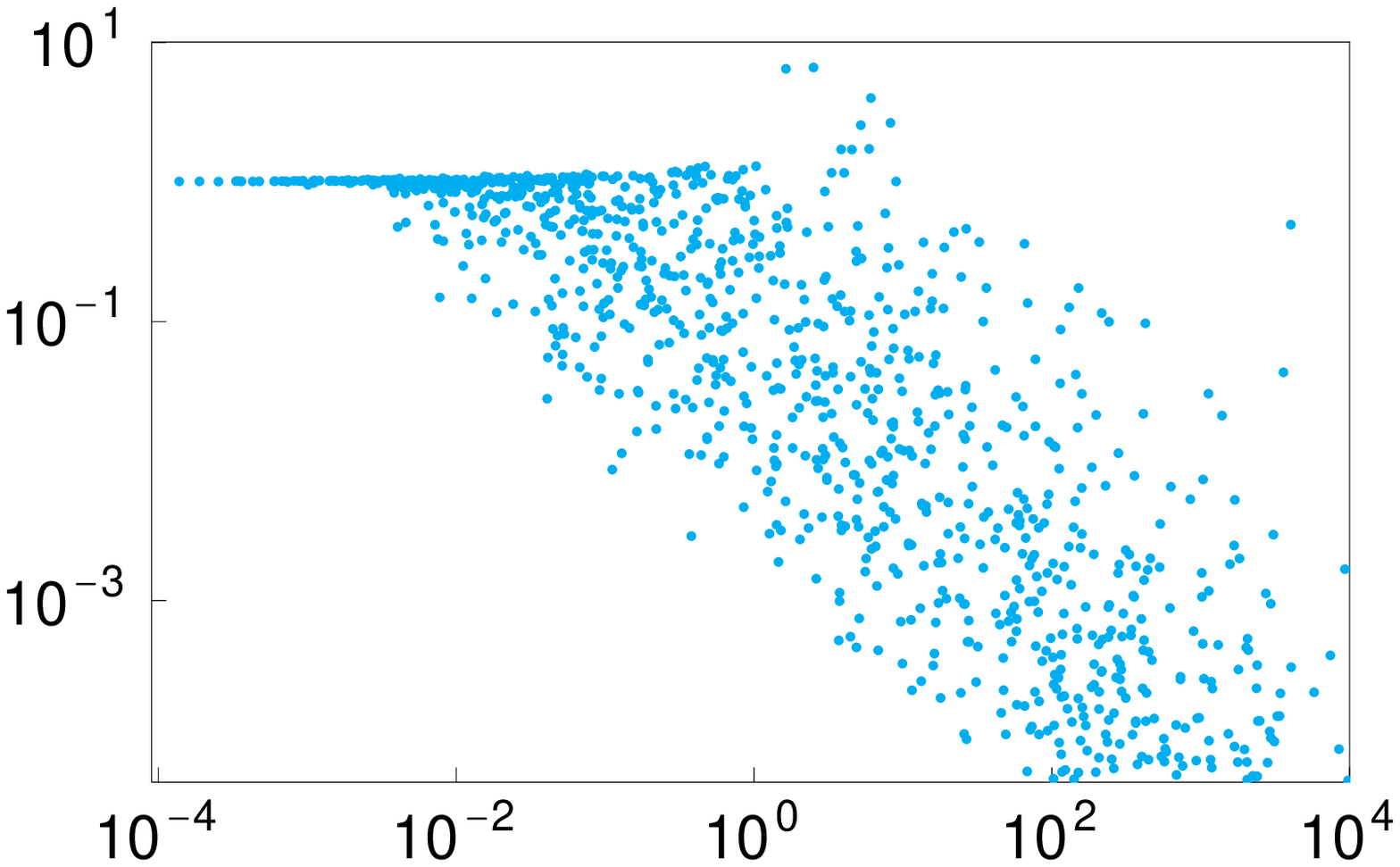}

\vskip-70mm
\noindent
{\rotate{$\langle m_{\nu}^{tree}\rangle / \langle m_{\nu}^{1-loop}\rangle$}

\vskip15mm
\hskip60mm
$\epsilon_e^2/\Lambda_e$  

\bigskip
\noindent
{\bf Figure 6: }{Calculated ratios of $\emass$ at tree-level to 
$\emass$  including the 1-loop 
corrections, as a function of $\epsilon_e^2/\Lambda_e$, see text.}

\section{Summary}

I have discussed the contributions of (bilinear) R-parity breaking
supersymmetry to neutrinoless double beta decay. 
$\znbb$ decay constrains only a subset of the possible bilinear
parameters, namely $\epsilon_e$ and the sneutrino VEV of the first
generation $\omega_e$. This is a general property of the theory and
does not mean any fine-tuning of parameters.  For the first generation
\rp parameters, on the other hand, $\znbb$ decay provides very
stringent limits, typically of the order of a few hundred keV up to a
few MeV. 

Interesting is the fact that $\znbb$ decay is strictly proportional to 
the alignment factor $\Lambda_e$ only at tree level. Once one goes 
to the next order in perturbation theory, $\znbb$ is sensitive to 
\rp parameters 
even in the case of perfect alignment. Although the limits in this 
special parameter range are less stringent than in the non-aligned 
case by about $3$ orders of magnitude, they nevertheless require that 
$\epsilon_e/\mu \le 0.01$, i.e. even for perfect alignment R-parity 
violation can not be maximal.

\bigskip

\centerline{\bf Acknowledgement}

I like to thank J.W.F. Valle for his always enthusiastic collaboration, 
S.G. Kovalenko for useful discussions and J.C. Rom\~ao for his 
help with the 1-loop calculations. This work was
supported by the Spanish DGICYT under grant PB95-1077 and by the
European Union's TMR program under grants ERBFMRXCT960090 and
ERBFMBICT983000.

\end{document}